# ULTRA-BROADBAND ELECTROMAGNETIC CONTROL USING A TRIPLE CIRCULAR RING METASURFACE: SURFACE WAVE PROPAGATION, BEAM STEERING, AND RCS REDUCTION

Md. Al Amin Chy[a, c], Tasnimul Hasan Tasneem[a, c], Sabbir Ahmad Chowdhury[a, c], and M.R.C. Mahdy[b,d,†]

[a]Department of Electrical and Electronic Engineering, Sylhet Engineering College, Sylhet, Bangladesh
[b]Department of Electrical and Computer Engineering, North South University, Dhaka, Bangladesh
[c]Mahdy Research Academy, Dhaka, Bangladesh
[d]NSU Optics Lab, North South University, Dhaka, Bangladesh

[†]Corresponding author: M.R.C. Mahdy, mahdy.chowdhury@northsouth.edu

## ABSTRACT

Traditional metasurfaces often face challenges in achieving simultaneous broadband functionality and dynamic adaptability, limiting their application in advanced electromagnetic systems. This paper introduces a cutting-edge metasurface featuring a triple circular ring design, engineered for multifunctional electromagnetic applications. This novel metasurface facilitates seamless strong propagation characteristics, enhanced beam shaping, and ultra-broadband radar cross-section (RCS) reduction using CST microwave studio. The proposed metasurface employs a cost-effective FR-4 substrate as its unit cell, demonstrating exceptional electromagnetic wave reflection properties. Across the high-frequency spectrum of 50–100 GHz, the metasurface exhibits amplitude and phase responses comparable to those of a conventional copper plate, except at approximately 91 GHz, where it achieves a significant dip in reflection. Our metasurface efficiently helps to propagate surface waves while maintaining substantial electric field and magnetic field amplitudes of nearly 1 V/m and 5×10^-3 A/m, respectively, contrasting starkly with a standard copper plate, which fails to sustain these fields over the entire surface. Moreover, it demonstrates notable radiation control by redirecting incident energy into a predefined direction (67° in the phi plane) while simultaneously minimizing radiation across a full 360° angular spectrum. Compared to a copper plate, which exhibits an unstable RCS ranging from -20 dB to 0 dB, our metasurface maintains a significantly reduced and stable monostatic RCS of -40 to -30 dB across an ultra-broadband range of 50 GHz. The results were validated through numerical simulations. By leveraging temporal boundaries, our approach establishes a novel framework for wavefront engineering, offering unprecedented scalability, efficiency, and adaptability. Bridging the gap between intelligent and passive metasurfaces, our findings hold immense potential for stealth technology, radar applications, and next-generation wireless communications. We believe that this work sets a novel way for metasurface engineering, paving the way for transformative advancements in electromagnetic science and practical implementations.

# 1 Introduction

The ability to control high-frequency electromagnetic waves is essential for numerous applications like energy harvesting[1], radar systems[2], wireless communications[3], and particle accelerators[4]. An engineered two-dimensional structure known as a metasurface has significantly advanced electromagnetic wave manipulation, which allows precise control over the phase & amplitude[5], and polarization[6]. To optimize waves, intelligent metasurfaces use deep learning, yet their actual scalability is limited by the size of their training datasets [7]. Comparably, at different angles, dual-band transmissive polarization converters show only a limited degree of stability [8]. Acoustic and elastic metasurfaces make mechanical wave manipulation possible, although their waveguide designs have a high structural stiffness [9]. The low-loss propagation efficiency of piezoelectric metasurfaces is also lacking, despite their potential in reconfigurable waveguides [10]. Compact devices cannot use metasurfaces combined with spin-polarized wavefronts because they consume excessive power despite achieving efficient energy conversion [11].

These constraints are addressed by our metasurface design, which maintains uniform wavefront propagation over wider frequency ranges, improves energy efficiency, and lowers loss. It is a superior option for surface wave engineering because it guarantees greater field confinement, propagation, and simpler manufacturing than current models. By adding temporal bounds, which offer an extra degree of control over waves and further broaden the potential of metasurfaces, these improvements are made achievable.

Even though these results set a new standard for temporal wavefront control, metasurfaces have a lot of potential for increasing electromagnetic radiation efficiency, especially in radiation control and directed beam shaping. Our approach provides a new approach of metasurface design by providing very high efficiency and directional radiation control with low reflections. At the total bandwidth, our metasurface produces more concentrated and sharper radiation, exceeding traditional designs in terms of energy economy and beam control.

These advancements in radiation management also support another important field of study: the lowering of radar cross-section (RCS), which is essential for applications involving electromagnetic wave suppression and stealth. These issues are resolved by the innovative metasurface that this work suggests. Using advanced phase control and broadband destructive interference, it consistently maintains lower RCS (-40 to -30 dB) over the 50–100 GHz range. With a reflection amplitude close to 1.0 and resilience to changing incidence angles, this metasurface avoids narrowband phenomena in contrast to resonance-based approaches. This metasurface is a game-changing solution in high-frequency stealth technology because of its ultra-broadband scalability, energy efficiency, and improved stealth characteristics.

Current methods provide some gains in energy efficiency, frequency conversion, and beam shaping. However, to the best of our knowledge, there isn't a single, scalable design that combines several sophisticated wave manipulation methods into one metasurface. To get beyond current constraints, we introduce a new reconfigurable metasurface that combines many cutting-edge features. In addition to improving stealth and reducing reflections, our approach achieves ultra-broadband RCS reduction (50–100 GHz). Highly directed radiation control improves beam shape, maximizing wireless sensing and communication.

By combining three innovative functionalities into a single unit cell, our designed metasurface redefines electromagnetic wave control: ultra-broadband RCS reduction, highly enhanced directed radiation, and enhanced surface wave propagation. The novel combination of spatial and temporal wave engineering allows for unparalleled scalability, flexibility, and efficiency. This work establishes a new standard in metasurface research by seamlessly connecting passive and intelligent metasurfaces, opening new possibilities in stealth applications, sophisticated radar technology, and next-generation wireless systems.

# 2 Background Studies

To unlock the new capabilities in dynamic wavefront engineering [12], while traditional passive metasurfaces rely on spatially varying geometries. These time-varying metasurfaces direct unprecedented flexibility in modulating electromagnetic waves, which allows for enhanced field confinement [13], frequency conversion [14], non-reciprocal transmission [15], etc. Metamaterial-inspired graphene and semiconductor structures have also enabled tunable broadband transmission and bandgap control [16].

Surface wave control has advanced through metasurfaces, ultrathin structures enabling precise tuning of amplitude, phase, and polarization. Since Chen et al. [17] introduced them in 2016, they've offered low-loss, compact alternatives to 3D metamaterials, despite challenges in tunability and large-area uniformity. Reconfigurable designs using liquid metals like Galinstan enable dynamic surface wave routing, as shown by Chu et al. [18]. Herbette et al. [19] further expanded this field with 3D metamaterials guiding HF-band waves via corrugated, negative-permittivity surfaces. RIS technologies now offer low-power, actively tunable surfaces for 6G systems [20], while THz metasurfaces achieve >80% cross-polarized reflection, enhancing broadband polarization control [21]. Surface waves are an essential tool for transporting energy and field localization. They are confined at the interface of a structured surface. Metasurface-assisted optical force engineering has also enabled sorting of plasmonic, dielectric, and chiral particles [22]. Engineering the permittivity and permeability of metasurfaces, which have been widely studied, can enhance surface wave propagation. Although specific control over surface wave propagation is possible with metasurfaces, issues including energy loss, limited bandwidth, and manufacturing complexity still exist. To enhance wave manipulation, dynamic, programmable, and hybrid metasurfaces have been investigated recently. Polarization-controlled metasurfaces have also been used for broadband optical pulling and tractor beam generation [23]. Every strategy, though, has drawbacks. While time-varying metasurfaces improve wavefront engineering, they also come with a higher price tag due to their complicated control systems [24]. Perovskite-based metasurfaces achieve effective terahertz wave control, although they deteriorate when exposed to environmental variations [25]. Although they incur larger losses at oblique angles, leaky-wave antennas increase control over wave dispersion [26]. Adaptive polarization conversion is made possible by programmable metasurfaces, although real-time adjustment of these surfaces requires a lot of processing power [27].

Metasurfaces enable precise electromagnetic wave manipulation, enhancing directional radiation while reducing unwanted reflections. Advanced directional radiation has had a major impact on heat management, wireless communication, and optical sensing. Compact, high-efficiency particle detectors [28] are advanced by surface Dyakonov–Cherenkov radiation, which is made possible by Dyakonov surface waves (DSWs) and enhances photon emission close to threshold velocities. $Si_3N_4$ metasurfaces provide directional emission control and strong field confinement, which are essential for accurate optical applications [29]. By exhibiting directional radiation and Kirchhoff's law violations, nonreciprocal metamaterials, like wire media in gyrotropic substrates, allow for dynamic heat control [30]. Simulations show nonlinear interactions and pump depletion in nanophotonics [31], where plasmonic particle-in-cavity nanoantennas (PIC-NAs) enhance second-harmonic generation (SHG) by centro symmetry breaking. Directional antennas and tailored beamwidth improve signal-to-interference ratio (SIR), range, and resolution for radar and wireless systems. These designs have been further optimized by boundary element models and stochastic geometry [32]. When combined, these advancements open new possibilities for tiny, highly efficient devices in the communication and photonics fields. Recent research has explored various metasurface designs for beam shaping in telecommunications and sensing. Antennas based on metasurfaces increase directivity [33], but they have trouble in dynamic environments. Although they lack wavelength flexibility, improved micro-LED designs provide better emission [34]. Beam steering is possible with tunable metasurfaces [35], but the power consumption goes up. Although they enhance beam shaping [36] Leaky-wave antennas have problems with thermal stability. High temperature metasurfaces are expensive but provide robustness [37]. Efficiency is improved by edge-effect corrections [38], though they are ineffective at wide frequencies. Nonreciprocal metasurfaces cause insertion loss but increase gain [39]. Although they are difficult to mass-produce, topological metasurfaces improve radiation efficiency [40]. Impedance mismatch is a problem with 5G metasurface antennas despite their increased bandwidth [41]. High frequencies are used by digitally controlled metasurfaces [42], although their calibration is intricate.

From the 1950s to the 1970s, early military research made clear how crucial it was to optimize material treatments and physical designs to reduce the radar cross-section (RCS) of airborne assets[43]. Research on reducing radar cross section (RCS) in the ultra-broadband frequency range of 50–100 GHz is crucial because it has applications in electromagnetic wave manipulation and stealth technologies. Several novel approaches have been investigated previously; however, the majority have encountered obstacles related to energy efficiency, bandwidth scalability, or practicality. Quantum-inspired metamaterial absorbers have also shown promising broadband absorption and stealth performance [44]. Su et al. (2018) showed that metamaterial tiles made with two different square ring unit cell designs reduced RCS by −10 dB throughout an ultra-wide frequency range of 6.2 GHz to 25.7 GHz [45]. While coding-based metasurfaces were effective in reducing RCS in the 7.5–17.5 GHz range, they were unable to maintain their efficacy at frequencies higher than 50 GHz [46]. At frequencies between 7.65 and 25 GHz, stereo coding meta-atoms offered polarization-independent designs with RCS reduction; nevertheless, their efficacy decreased [47]. According to [48], graphene-based metasurfaces demonstrated encouraging results with a 20 dB RCS reduction in the 10.3–19.2 GHz range. However, they were not appropriate for high-frequency applications due to energy dissipation at ultra-broadband levels. The usage of magnetic absorptive metasurfaces was restricted to ultra-broadband situations since they performed poorly beyond 30 GHz, although achieving significant reductions in the 3.4–18 GHz range [49]. Conventional metallic patterns had limited utility due to their high sensitivity to incidence angles and lack of broadband capabilities, although they are more affordable [50]. To achieve ultra-wideband scattering reduction (10.8–31.3 GHz), some designs used polarization-independent coding metasurfaces; however, these were unable to produce consistent reductions in higher-frequency ranges [51]. Wide-angle and polarization-independent characteristics were integrated with stereo meta-atom-based metasurfaces, which performed well in the 7.5–15.5 GHz range but were not scalable to ultra-high frequencies [52]. Although all-metal metasurfaces had broadband capability for frequencies ranging from 17 to 42 GHz, their performance instability and complexity of production were drawbacks [53]. While scalability, efficiency, and dynamic adaptability issues still exist, metasurface technology breakthroughs have shown great promise for electromagnetic wave manipulation, directional radiation control, and radar cross-section reduction, opening the door for further development and improvement.

# 3 Preliminary Concepts

## 3.1 Metasurfaces: Basic Principles

With recent progress, scientists have been able to create metasurfaces capable of performing multiple functions simultaneously. Their mathematical description relies on generalized sheet transition conditions (GSTCs), formulated from Maxwell's equations.

$$\nabla \cdot \mathbf{D} = \boldsymbol{\rho}, \qquad \nabla \cdot \mathbf{B} = \mathbf{0}, \quad \nabla \times \mathbf{E} - \frac{\partial \mathbf{B}}{\partial t}, \qquad \nabla \times \mathbf{H} = \mathbf{J} + \frac{\partial D}{\partial t},$$

When a metasurface is placed at the position z=0, the tangential components of the electric field and magnetic field are denoted as $\mathbf{E}_t$ and $\mathbf{H}_t$, respectively, must satisfy specific boundary conditions that relate their behavior on either side of the surface. These conditions emerge by integrating Maxwell's curl equations over the extremely thin region of the metasurface [54], [55]:

$$\hat{\mathbf{n}} \times (\mathbf{E}^+ - \mathbf{E}^-) = -j\omega\mu_0 \mathbf{M}_t, \qquad \hat{\mathbf{n}} \times (\mathbf{H}^+ - \mathbf{H}^-) = j\omega \mathbf{P}_t,$$

The quantities $\mathbf{E}^{\pm}$ and $\mathbf{H}^{\pm}$ indicate the electric and magnetic fields immediately above (+) and below (–) the metasurface, respectively. These equations form what are called the generalized sheet transition conditions (GSTCs), which provide a framework for understanding how metasurfaces manipulate electromagnetic waves via specially designed polarization effects [56], [57]. By adjusting the surface susceptibilities ($\overline{\overline{\chi_{ee}}}$ , $\overline{\overline{\chi_{mm}}}$ Reconfigurable metasurfaces achieve tunable control over the wavefront. Accordingly, the induced electric and magnetic polarizations are given by:

$$\boldsymbol{P_t} = \epsilon_0 \overline{\overline{\chi_{ee}}} \bullet \mathbf{E}_{avg}, \qquad \mathbf{M}_t = \overline{\overline{\chi_{mm}}} \bullet \mathbf{H}_{avg},$$

Researchers typically express the average field across the metasurface as $\boldsymbol{E_{avg} = E^+ + E^-/2}$. To achieve dynamic modulation of this field, one can integrate components like varactors, MEMS-based mechanisms, or materials that switch phase states [58], [59]. Such innovations make it possible to program the amplitude, phase, and polarization of waves, whether reflected or transmitted, in real time [60], [61].

## 3.2 Electromagnetic Wave Manipulation via Phase Discontinuities

If the phase profile *ϕ(x)* of a metasurface changes with position, the reflected wave acquires a corresponding phase gradient. To model this effect, we consider the metasurface to impose a surface impedance $\boldsymbol{Z_s\,(x)}$ that also varies spatially:

$$\Phi_r(x) = \tan^{-1}\left(\frac{Z_s\,(x) - \eta_0}{Z_s\,(x) + \eta_0}\right)$$

The direction of the reflected light follows the generalized Snell's law [62]:

$$k_{r,x} - k_{i,x} = \frac{\partial\phi(x)}{\partial x}, \quad \sin\theta_r = \sin\theta_i + \frac{\lambda}{2\pi}\frac{\partial\phi(x)}{\partial x}$$

This concept is fundamental to gradient metasurfaces, which allow for precise control over beam direction,

including steering and unusual reflection behaviors [63], [64]. A notable demonstration of this was provided by Yu and Capasso, who used metasurfaces with phase profiles that change across the surface [65], [66].

### 3.3 Directional Radiation and Wavefront Control

To manipulate wavefronts, metasurfaces create an aperture field $E_t(x)$, which determines the behavior of far-field radiation. The far-field is given by:

$$E_{far}(\theta) \propto \int_{-\infty}^{\infty} E_{far}(x) e^{-jk_0 x \sin\theta}\, dx$$

The far-field pattern is influenced by how the phase profile $\phi(x)$ is designed. For example, applying a linear phase shift $\phi(x) = \alpha x$ results in a beam pointing at an angle where $\sin\theta = \frac{\lambda\alpha}{2\pi[63]}$ [67]. Reconfigurable metasurfaces, such as leaky-wave and holographic variants, build on this concept to enable adaptive and dynamic beam steering [68].

### 3.4 Radar Cross Section (RCS) Reduction

Metasurfaces can lower the radar cross-section (RCS) by either absorbing incoming waves or scattering them in multiple directions. In absorptive approaches, lossy materials are used to convert electromagnetic energy into heat. The reflection coefficient is:

$$R(x) = \frac{Z_s - \eta_0}{Z_s + \eta_0}, \qquad where\ Z_s = R + jX$$

The reflection coefficient describes this behavior, and by properly adjusting it, perfect absorption ($|R| = 0$) can be realized [69]. Alternatively, coding metasurfaces that use phase elements like {0, π} can scatter waves through destructive interference:

$$E_s(\theta) \propto \sum_n R_n\, e^{-jk_0 x_n \sin\theta}$$

Using random or pseudo-random coding sequences helps suppress specular reflections, which has been effectively applied in broadband stealth technologies [70].

## 4 Unit Cell Configuration

The design concept of the proposed metasurface unit cell is illustrated in Figures 1. As shown in the top view of Figure 1(b), the structure features a triple-ring resonator configuration, purposefully developed to support multiple electromagnetic functionalities. For wave propagation analysis, nine identical unit cells are placed along the x-axis. The metasurface is constructed on an FR-4 substrate, an economical, flame-resistant, and glass-reinforced epoxy laminate chosen for its reliable dielectric properties. The total width of the structure is 6.102 mm, and a copper layer is applied on top of the substrate. All geometric parameters are fixed, except for the thickness of the copper layer, denoted as 'tc'. The structure is designed to be illuminated by a plane wave striking it head-on (normal incidence), resulting in highly directional emission and a noticeable reduction in radar cross-section (RCS). Functionally, the metasurface is built from three essential layers: a top resonator featuring three nested metallic rings, a central dielectric spacer, and a metallic ground plane at the base, which provides structural support. The geometric details of the unit cell, such as the

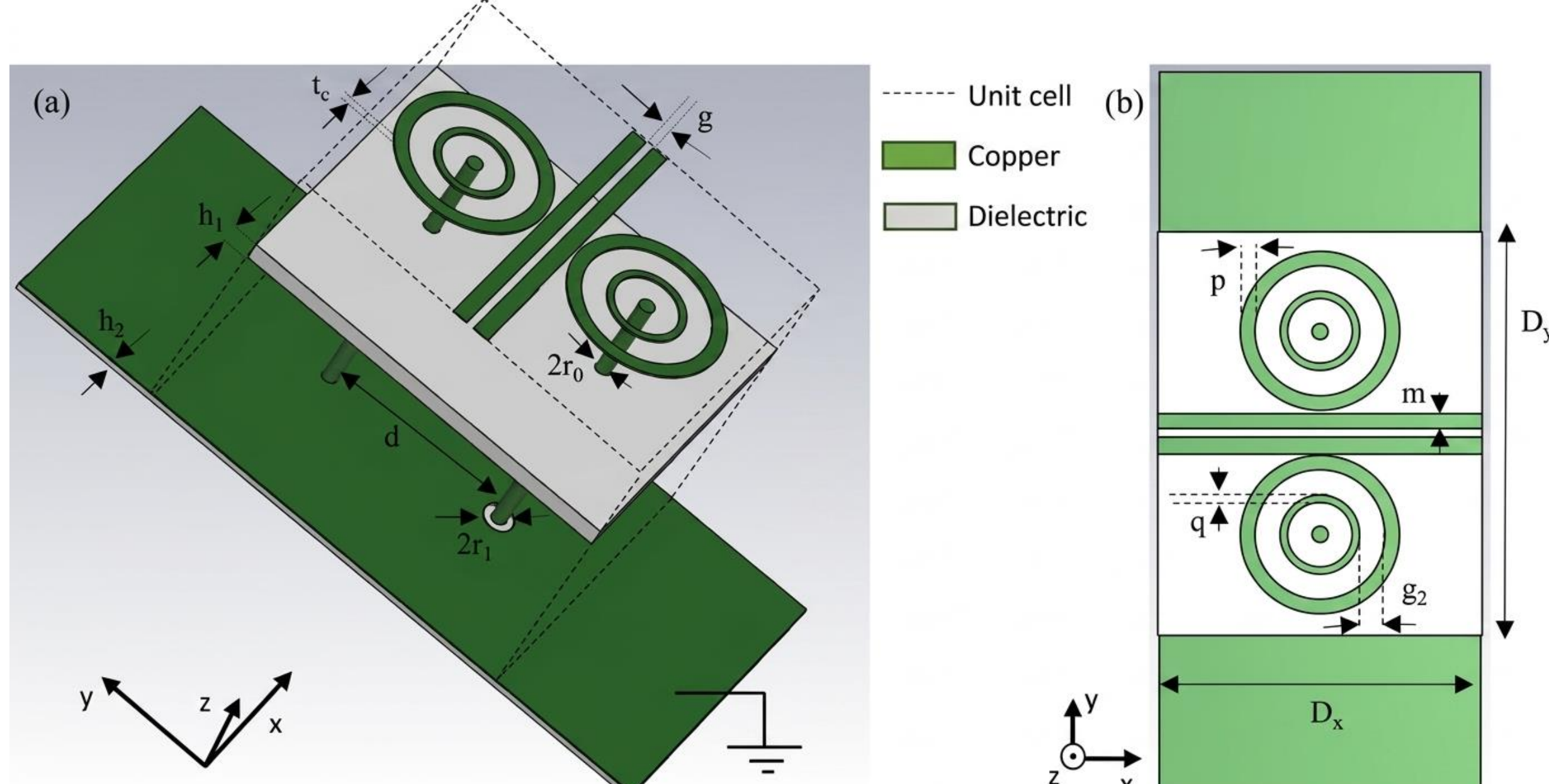

Figure 1 shows the unit cell design in different views. (a) displays a 3D view, where (b) illustrates a 2D view from top view, xy-plane. The dimensions are marked at their respective positions, where g = 0.0678 mm, $2r_0$ =0.1356 mm, $2r_1$ = 0.2712 mm, tc =0.00678 mm, $h_1$ = 0.16272 mm, $h_2$ = tc + 0.06102 mm (for dielectric), d = 1.55941 mm, Dy = 3.39 mm, Dx = 2.712 mm, p = 0.12204 mm, q = 0.06102 mm, m = 0.1356 mm, $g_2$ = 0.21018 mm.

substrate thickness (Dy) and the widths of the outer (p), middle (q), and inner ($2r_0$) rings, are labeled in Figure 1(b). These metallic rings are patterned onto an FR-4 substrate, a material with a relative permittivity (ε) of 4.3 F/m, magnetic permeability (μ) of 1 H/m, and a loss tangent of 0.025. The periodic distance between unit cells is fixed at T = 2.712 mm. For observing RCS performance and directional emission characteristics, a normally incident plane wave is used.

Further design details and simulation results for the complete unit cell are provided in Supplementary Note 2. To study the metasurface's electromagnetic response, simulations were carried out using the frequency-domain solver in CST Microwave Studio. The simulation environment was tailored for unit cell analysis, with periodic boundary conditions applied along the x and y directions. A Floquet port configuration was used, where the excitation was introduced along the z-axis through designated ports. To benchmark performance, a reference copper-based unit cell with the same geometric parameters (Dy and Dx) is also simulated. The reflection phase comparison between the proposed design and the copper reference shows a broadly similar trend, with a small phase shift around 91 GHz, as shown in Figure 9(b). Significant phase variations are observed at 70 GHz and 75 GHz, with shifts from -180° to 180°, respectively. Figure 9(a) displays the reflection magnitudes, revealing that the proposed unit cell maintains near-unity reflection across most of the spectrum, with a clear dip near 91 GHz.

Since metallic metasurfaces typically exhibit a reflection magnitude close to 1, minor variations in this value have little influence on RCS reduction [71]. To explore the metasurface's extended capabilities, specifically, its ability to support strong surface-wave propagation and enable temporal absorption, two distinct configurations were implemented. Electromagnetic Shielding: Perfect electric conductors (PECs) were positioned on both outer edges along the length of the metasurface to restrict wave propagation to the y-direction. Wave Excitation: A horn antenna was employed as the source of electromagnetic radiation, with further details outlined in Supplementary Note 1. The ring-shaped resonator offers improved stability due to its reduced sensitivity to changes in the angle of incidence and polarization of incoming waves. Its circular design also leads to more isotropic electromagnetic behavior when compared to square ring resonators, ensuring better uniformity across the structure. To analyze the electric field distribution, time-domain simulations were conducted using CST Microwave Studio. The simulation environment used open-space boundary conditions, and a discrete port was integrated within a horn antenna. The unit cells were arranged along both the x- and y-directions, while the horn antenna generated a y-polarized electric field that struck the metasurface at a 60° angle from the normal surface.

## 5 Results and Discussion

This section explores the three key functionalities of the proposed metasurface: broadband RCS reduction, highly directional radiation, and efficient surface wave manipulation with temporal modulation. We begin by analyzing surface wave behavior, highlighting enhanced field confinement and reduced propagation losses. Following that, we investigate how real-time control over electromagnetic energy is achieved through dynamic tuning of absorption via reactance modulation. We then focus on directional radiation, demonstrating improved beam steering and shaping capabilities. By carefully controlling the phase of reflected waves and leveraging destructive interference, the metasurface effectively minimizes monostatic radar cross-section (RCS) across a wide frequency band ranging from 50 to 100 GHz. This broad-spectrum suppression highlights its strong capability for stealth-oriented applications. Each of these functionalities and their broader impact on electromagnetic wave control are discussed in the sections that follow.

### 5.1 Strong Surface Wave Propagation and Temporal Absorption

In this analysis, we compare the behavior of surface wave propagation between a conventional copper surface (used as a reference) and our newly designed metasurface. A y-polarized electromagnetic wave is launched along the x-axis, traveling from the negative to the positive direction and striking the surface at a 60° angle relative to the z-axis. In this setup, the corresponding magnetic field is aligned along the z-direction. As illustrated in Fig. 2(a), the metasurface generates a significantly stronger Ey field component compared to the copper reference. This enhancement suggests improved energy confinement and more efficient propagation of surface waves.

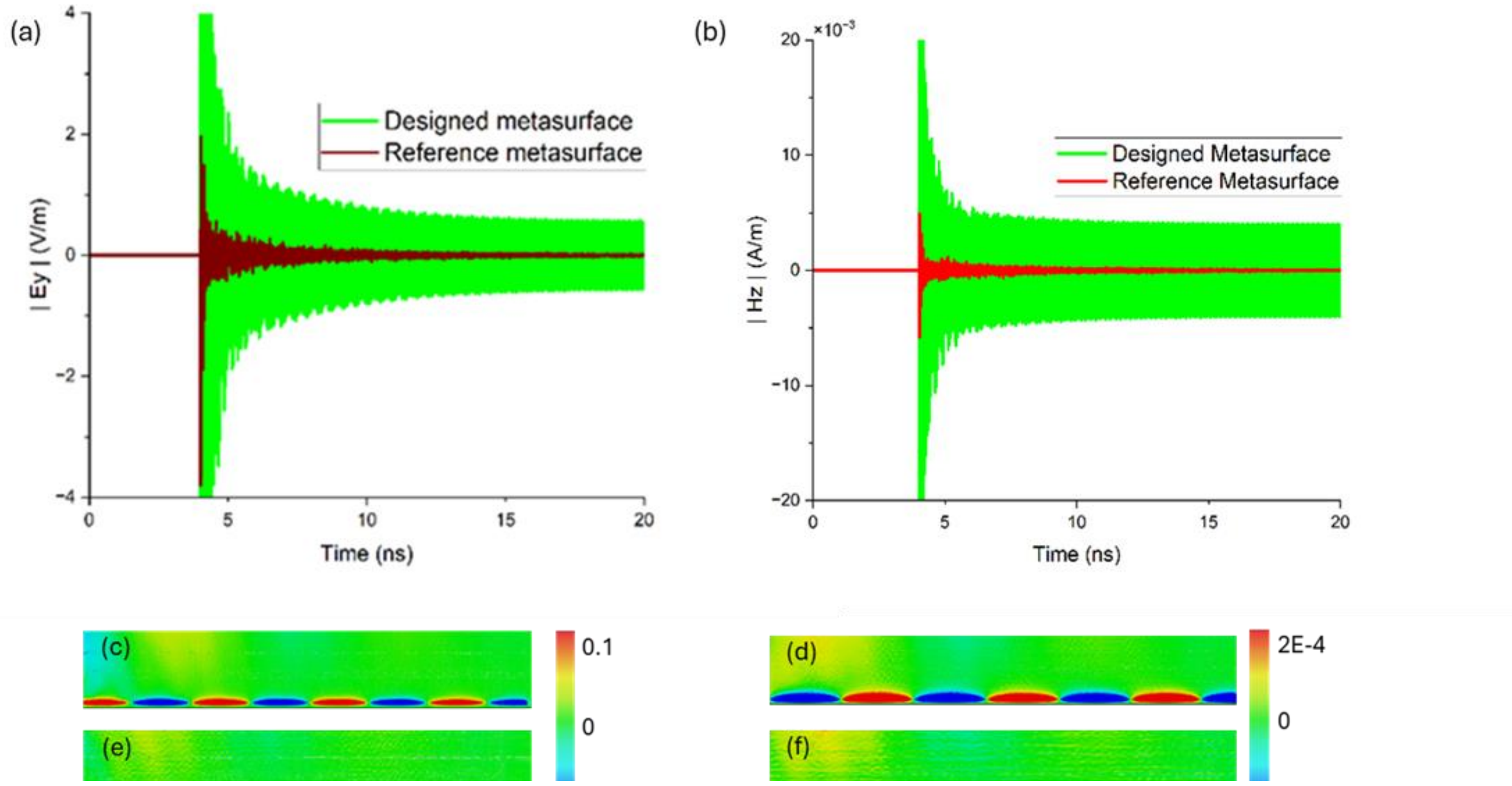

Figure 2 Illustrates how strong the electric field (Ey) and magnetic field (Hz) are distributed compared to the reference metasurface according to the time variation along the x-axis. (a) & (b) show 1D results of electric field and magnetic field located at the probing point at the illuminating side, marked as a black dot in Figs 5(d), of both metasurfaces. Additionally, 2D results evident a significant comparison of surface wave propagation at almost the ending moment (t = 19 ns) over the total 20 ns simulation time, where (c & e) for electric field and (d & f) for magnetic field distribution for the designed & reference metasurface, respectively.

On the other hand, the copper reference shows a much quicker drop-off in field strength, suggesting greater energy loss and a lower capacity to support surface-bound wave transmission. Similarly, Fig. 2(b) illustrates that our metasurface maintains a higher and more stable magnetic field, while the reference loses magnetic response more rapidly. This behavior implies that the designed structure facilitates more effective wave confinement, enabling stronger and more stable surface propagation. Further evidence can be seen in Figs. 2(c) and 2(e), where our metasurface preserves a well-formed and consistent wavefront, confirming its waveguiding efficiency. By contrast, the reference surface exhibits a dispersed and weakened field, which is a sign of higher energy dissipation. Additionally, the Hz field pattern in Fig. 2(d) for our design shows clear periodicity and structure, traits essential for controlling surface waves. In comparison, the Hz field for the copper surface in Fig. 2(f) appears irregular and poorly defined, reflecting its inability to guide waves effectively.

While copper is known for its excellent conductivity, its performance declines at high frequencies due to increased ohmic losses. These losses are affected by various parameters, including surface roughness[72], As frequency increases, surface resistance rises proportionally to the square root of frequency [73], and the thickness of the conductive layer [74]., which enhances resistive losses [75] and leads to faster attenuation of surface waves [76]. To further explain the energy behavior, the Poynting theorem is introduced, providing a foundation for analyzing electromagnetic energy conservation across the metasurface.

$$\nabla \cdot \mathbf{S} = -\frac{\partial u}{\partial t} - \mathbf{J} \bullet \mathbf{E} \tag{1}$$

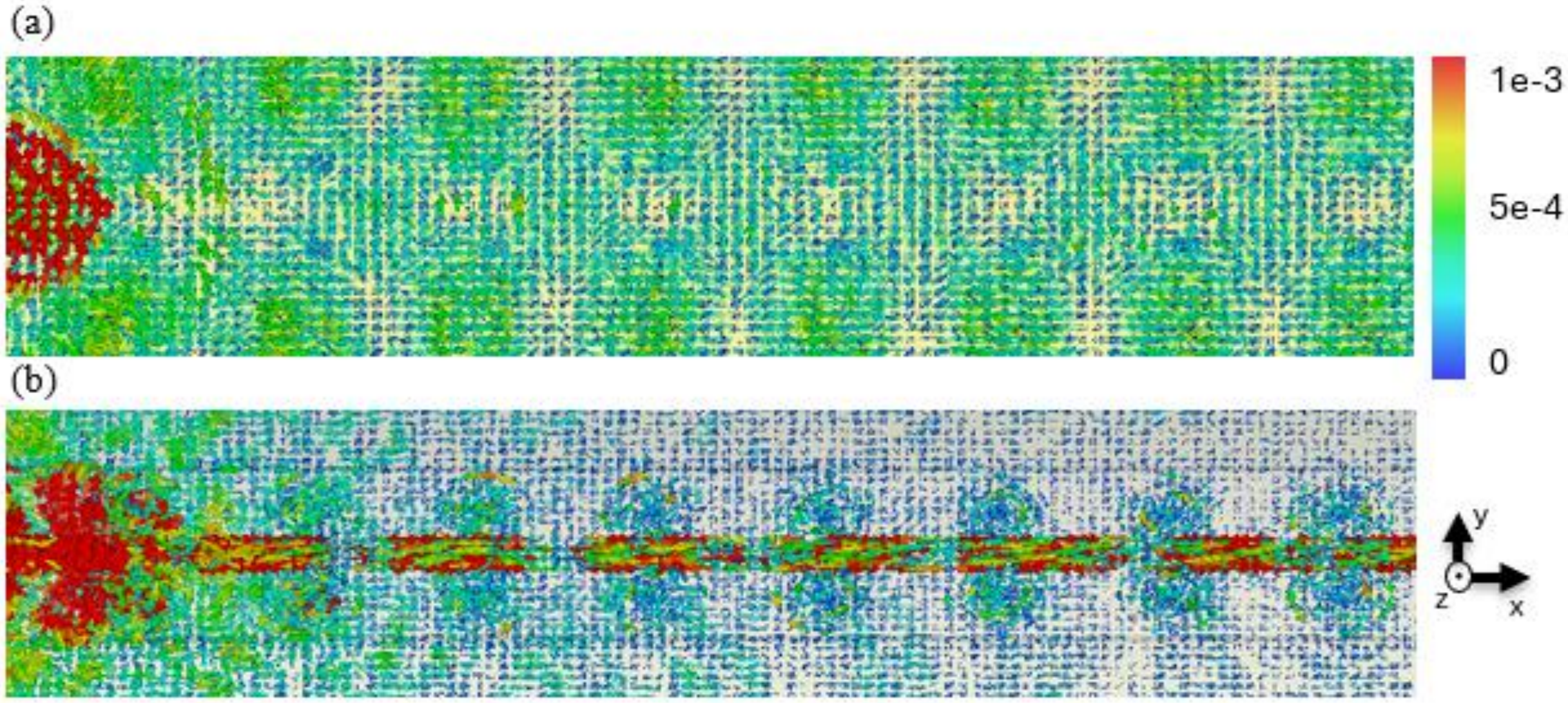

Figure 3: Illustrates the Surface current distribution (A/m) of both (a) the reference (copper) metasurface and (b) our designed metasurface at time t = 5 ns.

The directional flow of electromagnetic energy is captured by the time-averaged Poynting vector: $S = \frac{1}{2}\mathfrak{R}(\mathbf{E} \times \mathbf{H}^{*})$, where the symbols **E** and **H*** stand for the electric field and the complex conjugate of the magnetic field, respectively. The total electromagnetic energy stored per unit volume is denoted by u, and the rate of energy loss due to electrical conductivity is defined by the inner product **J·E**, with **J** as the current density. When this relation is applied to a thin conducting layer, as elaborated in [78], the total dissipated power is found by integrating across the surface area. For metasurface designs, the energy loss per unit length is caused by ohmic resistance, resulting from induced surface currents that can be described by the following formulation:

$$P_{loss} = \frac{1}{2}\,\mathrm{Real}\Big(\int_{Surface} J_S^{*}.E_s dS\Big) \tag{2}$$

The relationship between surface current density and surface conductivity σ is expressed as $J_S = \sigma E_S$ denotes the surface electric field. The equation highlights that ohmic losses on the metasurface result from the interaction between the induced surface current and the electric field, integrated over the surface. To visualize this, Figures 3(a) and 3(b) show the distribution of surface current $\mathrm{J_S}$ on the top layers of both the proposed metasurface and the copper reference at t = 5. As seen in Fig. 3(a), the designed metasurface exhibits concentrated surface currents primarily along two specific copper regions, while the rest of the surface shows relatively lower current levels, indicating spatially selective wave manipulation. This localized concentration suggests that energy is absorbed in a controlled and efficient manner, rather than being dissipated uniformly. Such behavior enhances dynamic surface wave confinement and directional energy control. In contrast, the conventional copper metasurface behaves passively, lacking the ability to modulate energy flow with the same precision.

In contrast, the reference metasurface exhibits a nearly uniform distribution of surface current across its entire structure, as shown in Fig. 3(a). This uniform current flow leads to consistent ohmic losses throughout the surface, as more of the incident electromagnetic energy is converted into heat. As a result, the reference metasurface demonstrates a higher overall absorption rate compared to the proposed design. However, this uniformity also indicates lower localized energy absorption, as the energy is not concentrated in specific regions. The absence of selective current pathways suggests that the reference metasurface lacks control over wave propagation and instead dissipates energy passively. On the other hand, the proposed metasurface effectively enhances surface wave propagation by confining both electric and magnetic fields and sustaining strong surface currents along designated areas. This localized and controlled energy interaction allows for improved wave manipulation. Consequently, the designed metasurface is more suitable for advanced electromagnetic applications such as beam shaping, stealth technology, and wireless energy transfer, where precise control over wave behavior is essential.

Furthermore, we extend our work by integrating engineered temporal discontinuities within the metasurface framework, and we achieve reversible conversion of propagating waves into non-radiating quasi-static magnetic modes. By adjusting the voltage-controlled switch, the flow of the surface wave can be halted before it reaches the farthest edge of the metasurface. In the proposed setup, the electric field is intentionally brought down to zero between 7 ns and 17 ns, even though the wave starts propagating from around 4 ns, as seen in Fig. 4(a). As illustrated in Fig. 5(a), the boundary reactance varies over time, initially exhibiting a capacitive characteristic, then becoming inductive, and finally returning to a capacitive state (C1 → L → C2). When the reactance becomes inductive, a static, uneven magnetic field is locked in place, and the electric field along the edge drops completely, causing the surface wave to vanish. Once the reactance returns to a capacitive state, the surface wave reappears and continues to flow. This effect is achieved by actively tuning the boundary conditions of the metasurface in real time. Initially, the incident TE-polarized wave, given by:

$$\mathrm{E}_i = \mathrm{E}_0 \sin(10\pi x + 10\pi\sqrt{3}z - 2\pi \times 30 \times 10^9 t)\,\mathbf{y} \tag{3}$$

propagates along the metasurface under a capacitive boundary condition ($C_0$). The corresponding magnetic field components are: $\mathrm{H}_i = \frac{\mathrm{E}_0}{\eta_0}\,(\frac{\sqrt{3}}{2}x - \frac{1}{2}z)\sin(10\pi x + 10\pi\sqrt{3}z - 2\pi \times 30 \times 10^9 t)$

For a transverse electric (TE) surface wave ($\beta,\omega_0$) propagating along a capacitive boundary with capacitance $C_1$, the electric and magnetic fields follow Equation (3). At t = 7 ns, shorting two metal copper plates via a voltage-controlled switch abruptly changes the boundary from capacitive to inductive, introducing an inductance L. This transition allows only transverse magnetic (TM) surface modes to exist at nonzero frequencies, replacing the TE mode. However, a TE mode can persist on an inductive boundary in free space if its frequency approaches zero ($\omega_1$=0). At t = $7^{+}$, the magnetic field component along the z-axis is given by: $\mathrm{H_z^{t=7+}} = \mathrm{H_0 e^{-j\beta x-\alpha_1 z}}\mathbf{z}$, while $H_0 = \frac{-E_0\beta}{\omega_0\mu_0}$, maintaining continuity in the magnetic field's z-component throughout the transition. Since $\omega_1$=0, the time-harmonic term $e^{j\omega_1 t}$ disappears, and the remaining field components simplify as:

$$H_y^{t=7+} = 0, H_x^{t=7+} = jH_0 e^{-j\beta x-\alpha_1 z}\mathbf{x}, \text{and } \mathbf{E}_x^{t=7+} = \mathbf{E}_y^{t=7+} = \mathbf{E}_z^{t=7+} = 0$$

This result confirms the presence of a static, spatially varying magnetic field extending along the x- and z-directions, while the electric field is entirely suppressed in all directions. The shift in boundary condition from capacitive to inductive is governed by a voltage-controlled switch, with the transition clearly illustrated in the 2D electric and magnetic field maps shown in Fig. 5(b–d). When the inductive state is established, the system enters what is often termed a "frozen eigenmode," satisfying the following relation:

$$j\omega_1 L_1.\left(\mathbf{n} \times \mathbf{H}_x^{t=7^+}\right) = E_y^{t=7^+} = 0$$

Instantaneous field behaviors along the metasurface are examined at various probe points, as indicated in Fig. 5(d). At Probe 1, which is located near the wave source, the electric field component Ey drops abruptly to zero, while the magnetic field Hz remains relatively steady, showing only minor variations.

This effect becomes particularly noticeable at higher frequencies around 30 GHz, clearly marking the moment when the electric field vanishes during the transition. The inductive response of the boundary is a key factor in governing the period of the electric field suppression, with both the intensity and duration of transient radiation being directly linked to the inductance. This behavior can be theoretically modeled using the Laplace transform method [77]. When the simulation reaches t = 17 ns, the boundary condition shifts back to a capacitive state, introducing a new surface capacitance $C_2$ defined by the eigenfrequency $\omega_2$ of the surface mode. This change leads to the reappearance of the electric field due to wave reflection and transmission, resuming normal propagation after the second transition. Out in free space, Maxwell's laws paint a simple picture: whenever the magnetic field twists through space, the electric displacement begins to change with time, and, in turn, the magnetic flux shifts whenever the electric field curls. For everything to behave properly, these fields also must flow seamlessly across any sudden jump in a surface's impedance, never breaking or "kinking" as they pass from one region to the next.

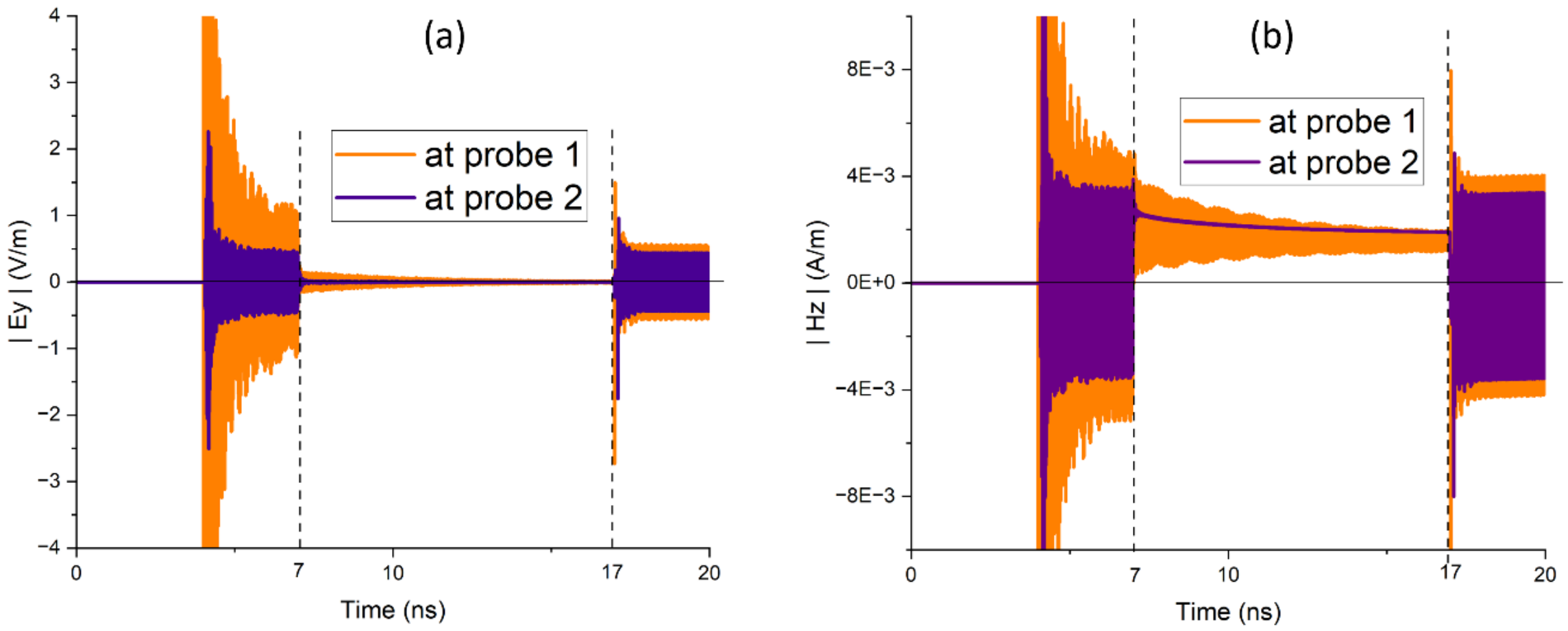

Figure 4: demonstrates the real value of two probes positioned at two different locations, probe 1 is placed near the illuminating side, and the farthest away position is occupied by probe 2, entirely the designed metasurface-directed Y-polarized electromagnetic waves including magnetic (Hz) and electric (Ey) field. (a) & (b) show the static and freezer fields of electric and magnetic, respectively, for the 7th to 17th time duration.

$$\mathbf{D}_{t=7^-} = \mathbf{D}_{t=7^+}, \mathbf{B}_{t=7^-} = \mathbf{B}_{t=7^+} \tag{4}$$

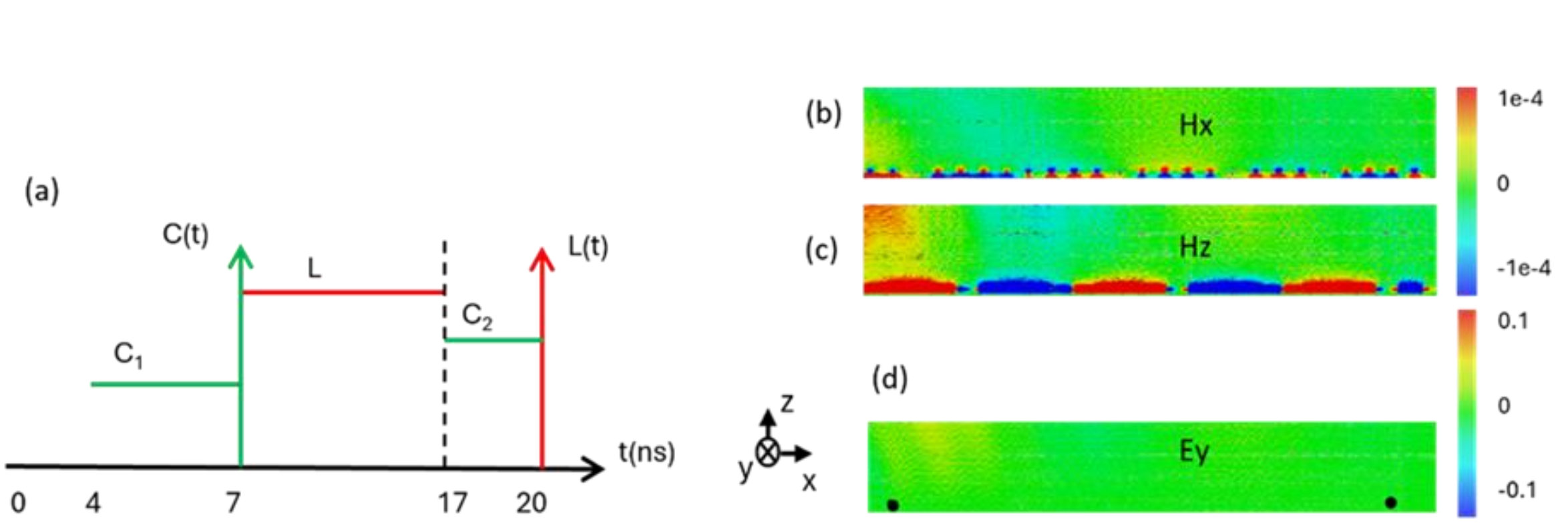

Figure 5:(a) illustrates how the boundary conditions evolve over the entire simulation timeframe. Figures (b)–(d) display the two-dimensional distributions of the electric and magnetic fields on the designed metasurface during the field-freezing interval (7 ns < t < 17 ns). In Figures (b) and (c), the magnetic field components Hx and Hz demonstrate a 90° phase difference, aligning with theoretical expectations. Furthermore, Figure (d) verifies the complete suppression of the electric field during the frozen state.

Using Equation (4), at t=17, i.e., $E_{ry}^{t=17^+} + E_{ty}^{t=17^+} = E_y^{t=17^-} = 0$ and $H_{rz}^{t=17^+} + H_{tz}^{t=17^+} = H_z^{t=17^-}$, the reflection and transmission coefficients after the second transition can be evaluated, as detailed in the reference [78]. According to the duality theorem, transitioning from an inductive to a capacitive boundary condition converts a TM-polarized electric field into a static surface wave. Reversing this transition restores the static field back into a propagating TM surface wave. This process mirrors behaviors observed in rapid plasma generation [79][80][81] and in magnetized plasmas subjected to fast-changing bias fields [82][83]. In this additional part, we demonstrate that an active surface wave can be momentarily suppressed, effectively halting the electric and magnetic fields. Once the second transition takes place, wave propagation resumes, reestablishing both electric and magnetic fields. This field localization enables active electromagnetic shielding and targeted jamming, validated by simulations and analytical field analysis under controlled reactance switching.

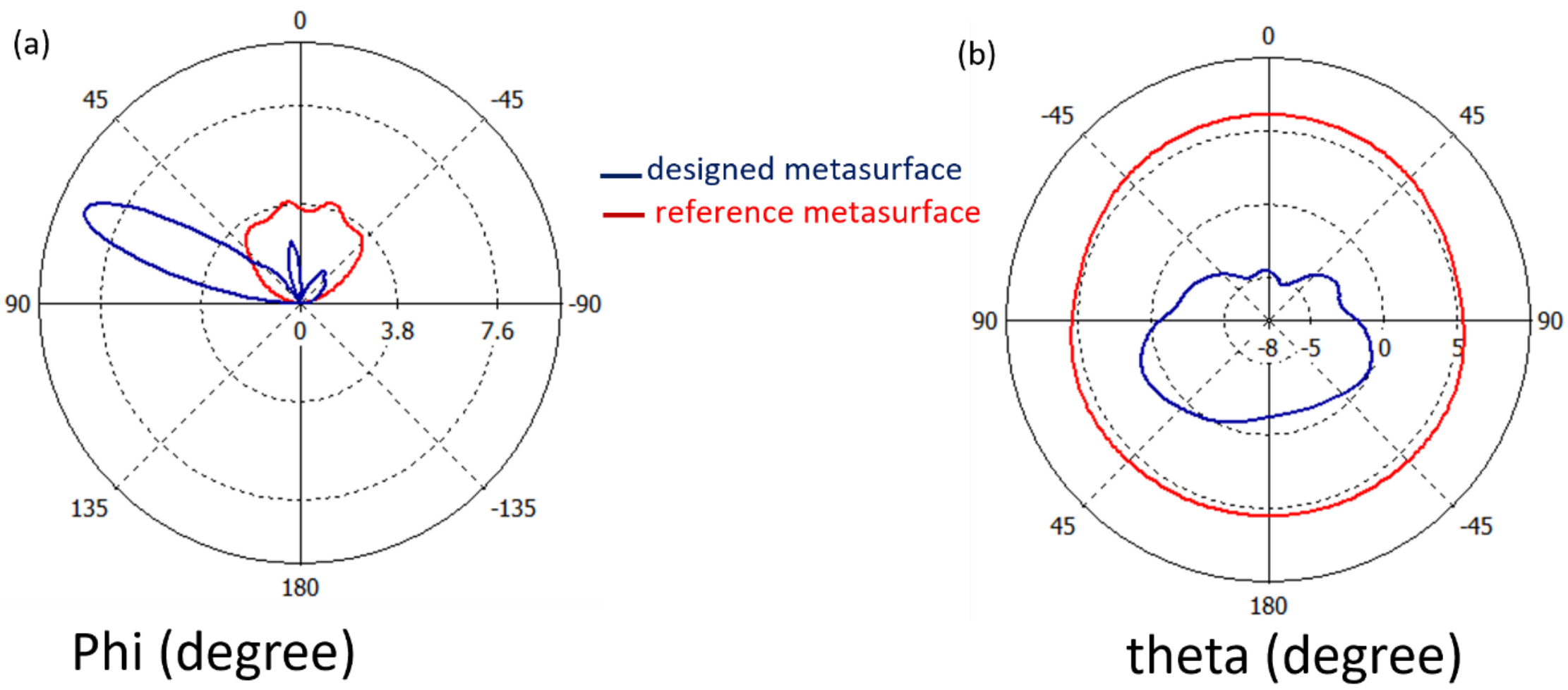

Figure 6 illustrates the far-field directivity results under normal plane wave incidence on the metasurfaces. (a) Shows a polar plot of the phi angle at 3.842 GHz, where the designed metasurface suppresses radiation in most directions while enhancing it in a particular direction compared to the reference surface. (b) Presents a polar plot of the theta angle at the same frequency, demonstrating that the designed metasurface significantly lowers overall radiation levels compared to the reference metasurface.

### 5.2 Enhanced Directional Radiation:

This work explores a newly developed metasurface structure aimed at improving directional radiation while suppressing undesired reflections. Far-field simulation results validate the metasurface's ability to achieve focused beam steering with minimal energy loss. As shown in Figures 6(a) and 6(b), the radiation patterns at

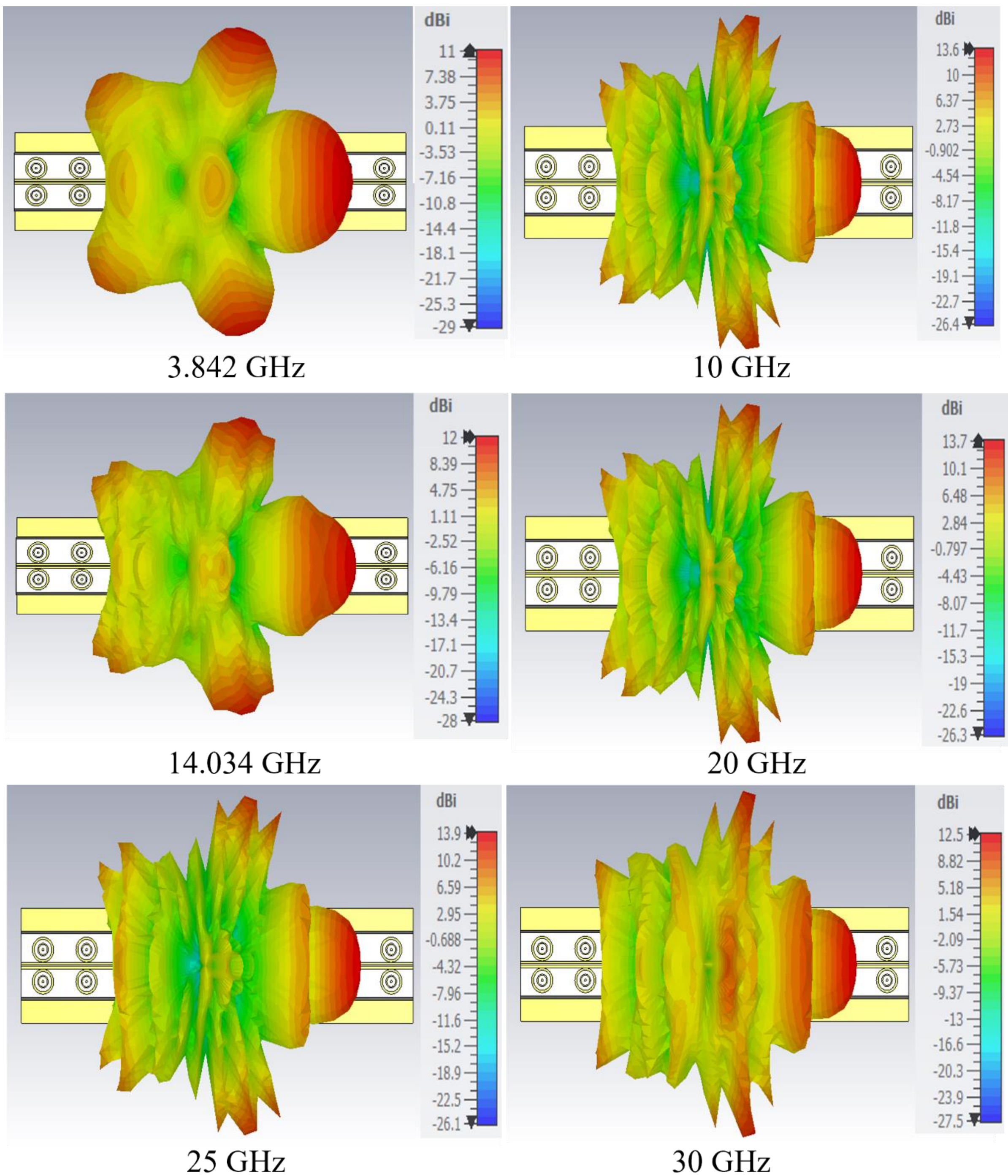

Figure 7 presents the 3D directivity pattern of the designed metasurface at different GHz, illustrating its ability to concentrate radiation in a specific direction while significantly suppressing it in other directions. This confirms the metasurface's directional radiation control across different frequencies.

3.842 GHz compare the performance of the proposed design against a conventional reference metasurface. Polar plots clearly reveal that the designed metasurface confines radiation to a targeted direction, significantly reducing emissions in other angles. In contrast, the reference metasurface displays a wider radiation spread, leading to higher power leakage into unintended directions. The directivity of the system, which quantifies this effect, is mathematically defined as:

$$D = \frac{4\pi\, U_{max}}{P_{rad}} \tag{5}$$

Where $U_{max}$ represents the maximum radiation intensity and $P_{rad}$ is the total radiated power. The increased directivity of the designed metasurface suggests that it efficiently redistributes radiated energy, concentrating it in the desired direction. The directivity, when represented on a decibel (dB) scale, is calculated using the formula:

$$D_{dB} = 10\, log_{10} D \tag{6}$$

Substitute Eq. (5) into Eq. (6) to define a dB scale, for two cases, the ratio $U_{max}/U_{avg}$ are 4.02 and 17.8 for the copper and the designed metasurface, respectively. Thus, the designed metasurface enhances directivity by a factor of 4.43 compared to the reference surface. Figure 7 presents the 3D radiation patterns at various frequencies in GHz for the designed metasurfaces. The reference metasurface shows a more isotropic radiation pattern, whereas the designed metasurface effectively enhances radiation in a targeted direction while suppressing it in others. The improved directivity is due to the ability of metasurfaces to engineer phase gradients [84], which control wavefront propagation. This is based on Generalized Snell's Law, which governs beam steering:

$$n_1 sin\theta_i - n_2 sin\theta_t = \frac{\lambda}{2\pi} \frac{d\Phi}{dx} \tag{7}$$

Where the refractive indices are represented by $n_1$ and $n_2$. The incident and transmitted angles are denoted by $\theta_i$ and $\theta_t$. And $\frac{d\Phi}{dx}$ is the induced phase gradient. The copper plate does not show any engineered phase gradient. The radiation is scattered broadly, leading to low directivity. $\theta_i = \theta_t$ which results in isotropic-like radiation, giving a directivity of only 6.03 dBi. A tailored phase gradient creates a controlled wavefront for the designed metasurface, enhancing beam steering. From Eq. (7), by carefully choosing $d\Phi/dx$ (phase variation per unit length), The metasurface can steer radiation into a specific direction, focusing energy and increasing directivity. Due to phase-controlled radiation, this enables a high directive beam, as seen in Fig. 7, which results in the maximum directivity of 13.9 dBi. This improvement represents a 4.4 times increase in radiation strength in the desired direction. The findings of this study suggest that the designed metasurface can be effectively utilized in antenna applications requiring precise beam steering and reduced backscattering, such as satellite communications and radar systems.

**5.3 Ultra-broadband RCS Reduction for high frequencies:**

Research in electromagnetic wave control and stealth applications often focuses on minimizing an object's radar cross-section (RCS), denoted as σ. A widely accepted formula represents the RCS based on far-field conditions [85]:

$$\sigma = 10 \log \left[ \lim_{r\to\infty} 4\pi r^2 \frac{|E^S|^2}{|E^I|^2} \right]^2 \tag{8}$$

In this formula, $|E^S|$ and $|E^I|$ refers to the magnitudes of the scattered and incident electric fields, respectively. For effective wave absorption, the metasurface must match the impedance of free space. The absorption rate of the proposed metasurface can be derived using scattering parameters that vary with frequency [86][87]:

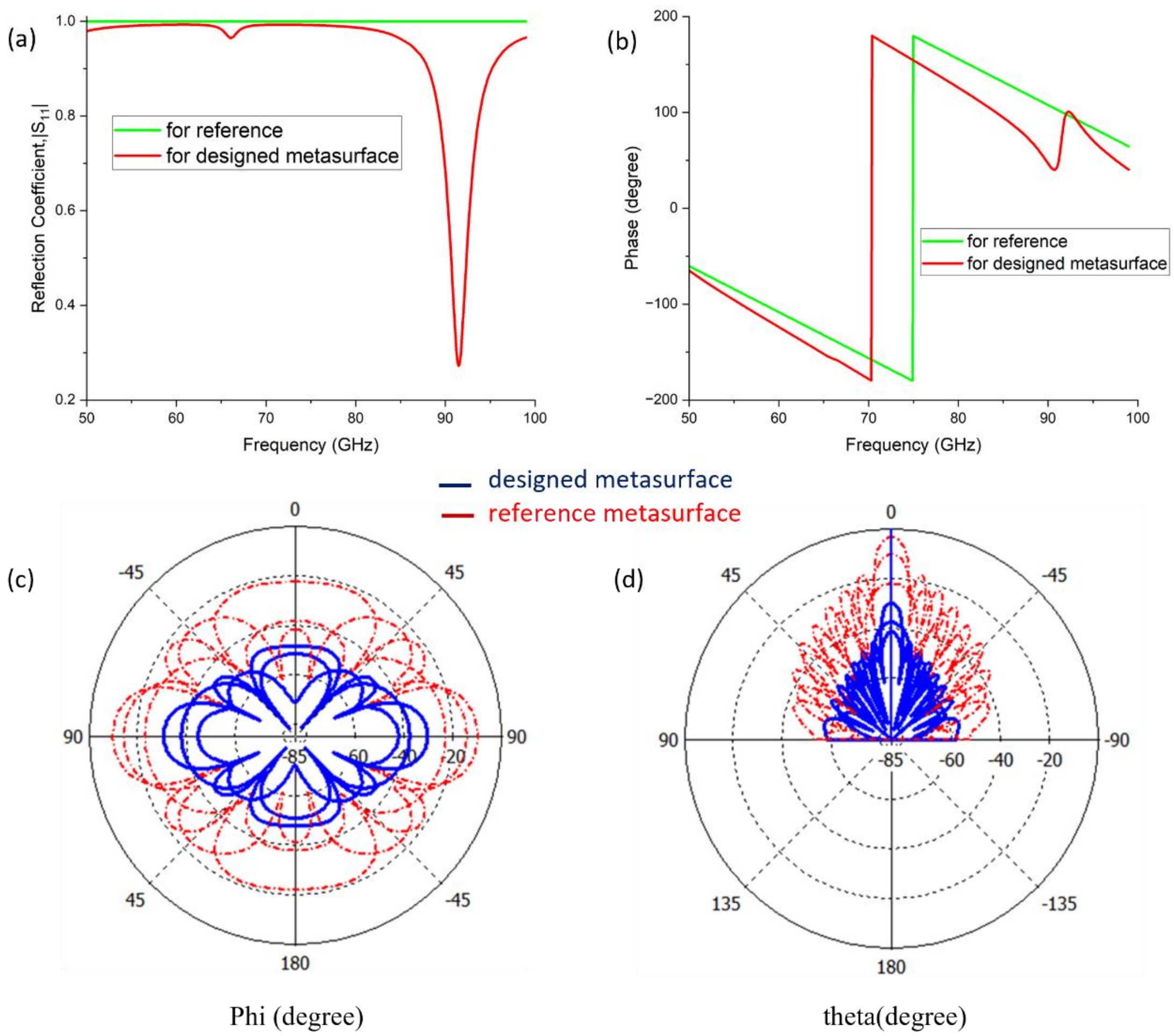

Figure 8: Farfield results. (a) and (b) show the reflection amplitude and phase of the unit cell, respectively. (c) and (d) display polar graphs of bistatic scattering in the range of -85 to 0 dB at frequencies like 50, 78, and 99 GHz. From both observation of azimuthal(phi) and elevation(theta) angle, this proves that the bistatic scattering is reduced significantly in all directions for the designed metasurface compared to the reference (copper) surface.

$$A(\omega) = 1 - |S_{11}(\omega)|^2 - |S_{21}(\omega)|^2 \tag{9}$$

Here, $S_{11}(\omega)$ and $S_{21}(\omega)$ are the reflection and transmission coefficients. However, since a layer of annealed copper is placed behind the metasurface, transmission is effectively eliminated, simplifying the equation as follows:

$$A(\omega) = 1 - |S_{11}(\omega)|^2 \tag{10}$$

This simplified expression shows that only the reflection coefficient influences the absorption rate lower reflection coefficient directly leads to higher absorption. Such metasurfaces are valuable in reducing electromagnetic interference in sensitive devices, communications systems, and other electronics by effectively absorbing unwanted signals. To investigate the radar cross-section (RCS) properties of the developed structure, simulations were conducted using the time-domain solver available in CST Microwave Studio. In these simulations, a plane wave was introduced through a hexahedral mesh setup. To gain deeper insights into the structure's reflective characteristics, additional simulations were performed using the frequency-domain solver. This setup incorporated periodic boundary conditions along the x and y directions, while a Floquet port was assigned along the z-axis to facilitate excitation with both transverse electric (TE) and transverse magnetic (TM) wave modes.
A detailed analysis of the RCS performance revealed that the proposed metasurface demonstrates a significant reduction in monostatic RCS. As shown in Fig. 9, the metasurface outperforms a benchmark reference made from solid copper with the same physical dimensions. In the frequency range between 50 GHz and 100 GHz, the monostatic RCS values generally range from –40 dB to –30 dB when measured at normal incidence.
Notably, the scattering behavior of the designed metasurface contrasts strongly with that of the copper reference surface, following almost inverse trends across the observed spectrum. This contrast in behavior can be attributed to the metasurface's ability to control the phase of the reflected electromagnetic waves. As a result, it achieves lower RCS values than the reference design. Figure 9 confirms that the designed metasurface significantly reduces RCS over a wide 50 GHz bandwidth when compared to the copper surface. This performance is achieved despite the reflection amplitude of the metasurface unit cell remaining close to 1.0 across most of the frequency band, as shown in Fig. 8(a), except for a sharp drop observed around 91 GHz. Additionally, Fig. 8(b) indicates that the reflected phase response closely follows that of the reference surface, except for a noticeable phase shift near 91 GHz. These findings imply that the metasurface utilizes broadband destructive interference and tailored phase responses to suppress backscattering. Since the total scattered electric field ($E^S$) is the cumulative result of reflections from all unit cells, the monostatic RCS (σ) can be expressed as:

$$\sigma \propto \left| \sum_{n=1}^{N} R_n e^{j\Phi_n} e^{-jk.r_n} \right|^2$$

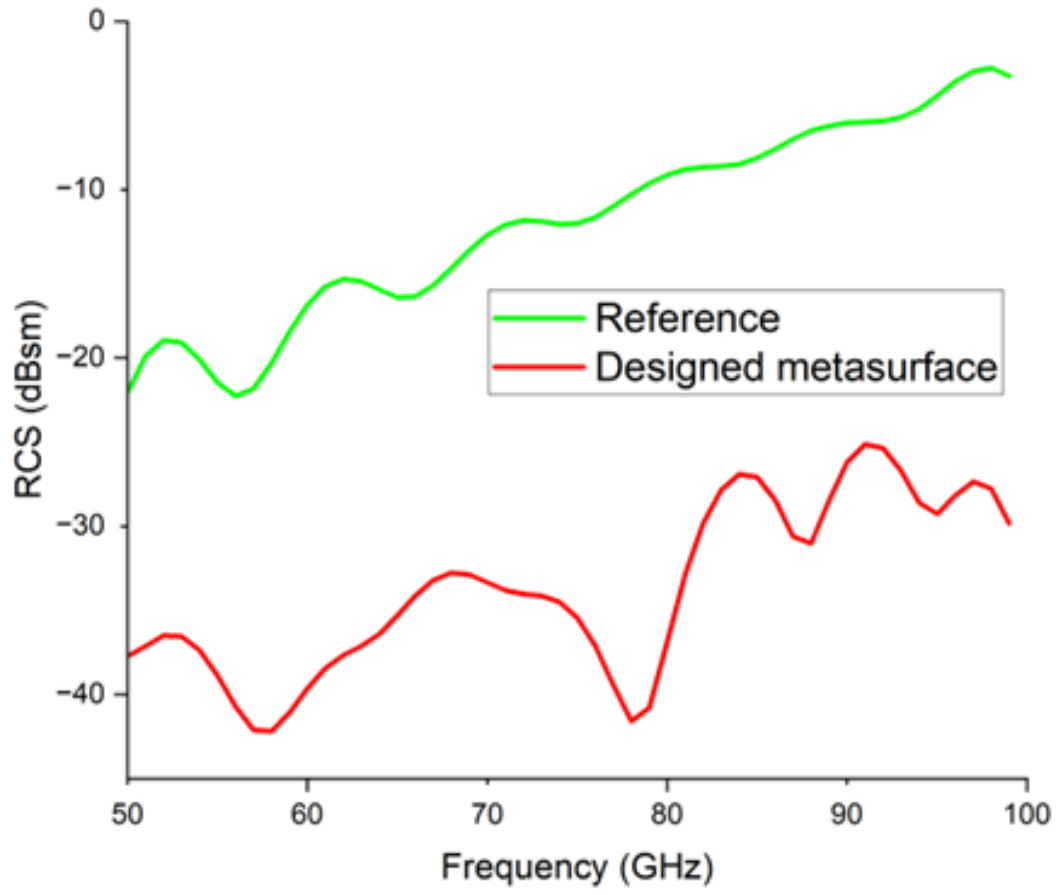

Figure 9: shows a monostatic RCS graph of the designed metasurface, which is significantly lower compared to the reference (copper) surface with such a wide bandwidth of 50 GHz.

Here, $R_n$ and $\Phi_n$ are the reflection amplitude and phase of the nth unit cell, $k.r_n$ represents the spatial phase delay, and N is the total number of unit cells. For a conventional metallic surface with high reflectivity ($R_n \approx 1$) and little to no phase variation ($\Phi_n \approx 0$), the reflected waves combine constructively, resulting in strong backscatter and a high RCS. In contrast, if the metasurface is engineered such that the reflected waves interfere destructively:

$$\sum_{n} R_n e^{j\Phi_n} \approx 0$$

then the RCS is significantly reduced. Even when the reflection magnitude remains high, a metasurface with precise phase engineering can redirect the reflected energy away from the incident direction, effectively lowering the RCS [88]. By incorporating a spatially varying phase profile, the metasurface introduces a phase gradient defined as [89]:

$$\Phi_n = \alpha n$$

where α represents the constant phase increment per unit cell. This gradient redistributes the reflected energy into multiple directions rather than concentrating it in the backscattering path. The phase profile of the designed metasurface remains nearly identical to that of the reference surface, except at around 91 GHz, where a significant phase shift occurs. This localized shift helps steer the reflected waves away from the source direction, supporting RCS suppression, an approach that aligns with findings from earlier metasurface-based scattering control studies [90]. The intensity of the backscattered field is mathematically defined as [91]:

$$\sigma_{mono} = \sum_{m,n} R_m R_n e^{j(\Phi_m - \Phi_n)}$$

Out-of-phase reflections from individual unit cells interfere destructively, effectively lowering the overall backscattered signal. Even slight phase shifts in the range of approximately 10° to 30° can have a significant impact on how the scattered waves are distributed across different angles, leading to a noticeable reduction in monostatic RCS. Notably, at around 91 GHz, the designed metasurface shows a sharp dip in reflection amplitude. This suggests that, at this frequency, the metasurface achieves localized impedance matching, meaning its surface impedance closely aligns with that of free space.

$$Z_S = \sqrt{\frac{\mu_0}{\epsilon_0}} = Z_0$$

At this point, $Z_S$ represents the surface impedance of the metasurface, while $\mu_0$ and $\epsilon_0$ denote the magnetic permeability and electric permittivity of free space, respectively. $Z_0$ refers to the intrinsic impedance of free space [92]. At this frequency, the structure behaves like an efficient absorber, almost eliminating reflections and thus lowering the RCS. Outside that narrow band, the metasurface maintains wideband RCS suppression primarily through precise phase control rather than a simple reduction in reflection amplitude. Broadband scattering mitigation is achieved through several synergistic mechanisms: (i) destructive interference spanning a wide frequency range, (ii) multi-resonant phase gradients, (iii) localized impedance matching near 91 GHz, and (iv) scattering-cancellation effects that collectively minimize coherent backscatter across the spectrum. These findings underscore phase-driven approaches as more effective for RCS reduction than traditional amplitude-based methods. By tailoring the phase delay imparted by each unit cell, the metasurface redirects scattered energy away from the backscattering direction, consistent with earlier studies on phase-gradient designs [93]. Each unit cell is tuned to foster destructive interference in the monostatic direction, dramatically reducing bistatic scattering. Consequently, the total scattered field can be regarded as the vector sum of these phase-shifted contributions, resulting in a marked drop in overall backscattered power.

$$E_{SC} = \sum_{n=1}^{N} a_n e^{j(\Phi_n + \beta_n)}$$

Here, $a_n$ represents the amplitude, $\Phi_n$ is the phase shift of the n-th unit cell, and $\beta_n$ denotes the applied phase gradient. By carefully selecting $\beta_n$, destructive interference can be induced in the backscattering direction, leading to a significant reduction in radar cross-section (RCS). This strategy is consistent with the conclusions drawn in [94]. Numerical simulations further validate that the proposed metasurface effectively lowers bistatic scattering across the examined frequency range. Polar plots of the scattering pattern indicate a notable decrease in scattered power across both azimuthal (phi) and elevation (theta) angles, confirming that the metasurface redirects energy away from the direction of incidence, as previously observed in [95]. This capability to suppress bistatic scattering across a wide frequency band holds strong promises for advancements in stealth technologies,

RCS minimization, and electromagnetic wave control. The proposed metasurface illustrates its potential for real-world applications such as radar evasion, improved wireless communication, and mitigation of electromagnetic interference.

# 6 Conclusion

This research introduces an innovative metasurface design featuring triple circular ring architecture, which marks a notable step forward in electromagnetic wave manipulation. The configuration is engineered to support multiple advanced functionalities, including efficient surface wave transmission with temporal absorption, directional beam control, and substantial ultra-broadband RCS reduction. All these capabilities are realized on an FR-4 substrate, enhancing both performance and practical integration. Compared to conventional metallic surfaces, the proposed metasurface exhibits enhanced electric and magnetic field confinement, enabling superior energy localization. A distinguishing feature of the structure is its dynamic surface reactance modulation, which allows for adaptive electromagnetic behavior within a defined temporal framework. Furthermore, the metasurface demonstrates effective redirection of scattered energy, resulting in consistently low RCS values across a wide frequency range. Due to its distinctive features, the proposed metasurface is well-suited for use in electromagnetic stealth technologies, advanced radar systems, and wavefront control applications. By bridging the gap between conventional passive surfaces and reconfigurable intelligent metasurfaces, this design establishes a promising foundation for future developments in stealth technology, advanced radar systems, and next-generation wireless communication platforms.

# Credential Authorship Contribution Statement

**Md. Al Amin Chy**: Led conceptualization, methodology, investigation, data curation, software, visualization, and formal analysis; wrote and revised the manuscript. **Tasnimul Hasan Tasneem**: Contributed to formal analysis, resources, and visualization. **Sabbir Ahmad Chowdhury**: Supported formal analysis, resources, and visualization. **M.R.C. Mahdy**: Provided supervision, project administration, investigation, resources, and visualization.

# Competing Interests

The authors confirm that they have no financial or personal relationships that could have inappropriately influenced the research and its findings.